\documentclass[pdflatex,sn-standardnature]{sn-jnl}

\jyear{2021}%

\theoremstyle{thmstyleone}%
\theoremstyle{thmstyletwo}%

\theoremstyle{thmstylethree}%

\begin{document}

\title[Article Title]{Designing High-$T_C$ Superconductors with BCS-inspired Screening, Density Functional Theory and Deep-learning }


\author*[1,2]{\fnm{Kamal} \sur{Choudhary}}\email{kamal.choudhary@nist.gov}

\author[1]{\fnm{Kevin} \sur{Garrity}}

\affil*[1]{\orgdiv{ Material Measurement Laboratory, Maryland 20899, USA }, \orgname{National Institute of Standards and Technology}, \orgaddress{ \city{Gaithersburg}, \postcode{20899}, \state{MD}, \country{USA}}}

\affil[2]{\orgname{Theiss Research}, \orgaddress{ \city{La Jolla}, \postcode{92037}, \state{CA}, \country{USA}}}


\abstract{We develop a multi-step workflow for the discovery of conventional superconductors, starting with a Bardeen–Cooper–Schrieffer inspired pre-screening of 1736 materials with high Debye temperature and electronic density of states. Next, we perform electron-phonon coupling calculations for 1058 of them to establish a large and systematic database of BCS superconducting properties. Using the McMillan-Allen-Dynes formula, we identify 105 dynamically stable materials with transition temperatures, $T_C$ $\geq$ 5 K. Additionally, we analyze trends in our dataset and individual materials including MoN, VC, VTe, KB$_6$, Ru$_3$NbC, V$_3$Pt, ScN, LaN$_2$, RuO$_2$, and TaC. We demonstrate that deep-learning(DL) models can predict superconductor properties faster than direct first principles computations. Notably, we find that by predicting the Eliashberg function as an intermediate quantity, we can improve model performance versus a direct DL prediction of $T_C$. We apply the trained models on the crystallographic open database and pre-screen candidates for further DFT calculations.}




\maketitle

\section{Introduction}

Since the discovery of superconductivity in 1911 by Onnes \cite{kamerlingh1911resistance}, the identification of novel superconducting materials, especially those with high transition temperatures ($T_C$), has been an active area of research in condensed matter physics \cite{poole2013superconductivity,rogalla2011100}. The highest temperature conventional superconductor in ambient conditions, MgB$_2$ ($T_C$ = 39 K), was discovered relatively recently\cite{nagamatsu2001superconductivity}, and progress in high pressure hydrogen-based superconductors\cite{pickard2020superconducting} and topological superconductors\cite{sato2017topological} further motivates the search for additional ambient or near-ambient condition superconductors with high $T_C$ that may be suitable for scientific or industrial applications.

There have been several previous efforts to computationally and/or experimentally discover superconducting materials falling into certain materials classes, such as transition metals \cite{roberts1976survey}, A15 , B1 \cite{kihlstrom1985tunneling,stewart2015superconductivity}, AB$_2$ compounds \cite{ivanovskii2003band,buzea2001review,nagamatsu2001superconductivity}, cuprates \cite{plakida2010high}, iron-based compounds \cite{hosono2015iron}, hydrides \cite{shipley2020stability,liu2017potential,drozdov2015conventional} and many other material-classes \cite{poole2013superconductivity,shipley2021high,yuan2019recent,nagamatsu2001superconductivity,rodriguez1990optical,subedi2013electron,duan2019ab,kolmogorov2010new,gao2010high,duan2014pressure}. However, a more systematic data-driven search should help expedite the discovery of potentially high-$T_C$ superconductors. 
Moreover, machine learning has become popular in the search for superconductors. There have been several reports of machine-learning applications for finding superconductors\cite{hutcheon2020predicting,stanev2018machine,yuan2019recent,xie2022machine}, but thus far they are mostly based on chemical formulas, and lack detailed atomic structure information that can be critical for superconducting behavior.

Two key ingredients required to computationally identify Bardeen–Cooper–Schrieffer (BCS) conventional superconductors \cite{cooper2010bcs, giustino2017electron} with high-$T_C$ are: 1) a robust computational workflow, and 2) a database of curated materials with prior knowledge such as elastic constants and electronic density of states. Using density functional theory perturbation theory (DFT-PT), electron-phonon coupling (EPC) can be calculated and used to predict $T_C$ with reasonable accuracy for arbitrary materials\cite{giustino2017electron, PhysRevB.101.134511}. However, the computational expense of these calculations is very high when compared to a single DFT self-consistent energy calculation, especially when fully converging the relevant $k$-point and $q$-point Brillioun zone samplings for electrons and phonons. Hence, a fast and reliable computational workflow for identifying BCS type conventional superconductors that balances computational cost, accuracy, and scope is needed.

In this work, we develop such a computational approach to discover BCS superconductors, combining several methods at various levels of computational expense and accuracy. We start with a BCS-inspired pre-screening based for materials with high Debye temperature ($\theta_D$) and high electron density of states (DOS) at Fermi-level ($N(0)$), using the existing JARVIS-DFT database \cite{choudhary2020joint}. We then develop and apply a DFT-PT workflow to compute $T_C$ using electron-phonon coupling and the McMillan-Allen-Dynes formula \cite{mcmillan1968transition}, with initially low convergence settings. We benchmark the DFT workflow on known superconductors and apply it to materials from our pre-screening step. For the best candidates, we perform additional convergence tests to validate our predictions. In addition, we use the dataset to develop deep-learning models using the atomistic line-graph graph neural network (ALIGNN)  \cite{choudhary2021atomistic,choudhary2021recent} to predict the Debye temperature, electronic DOS, $T_C$, and electron-phonon coupling parameters for arbitrary crystal structures.

We utilize the publicly available JARVIS \cite{choudhary2020joint} infrastructure to achieve the goals mentioned above. JARVIS (Joint Automated Repository for Various Integrated Simulations, \url{https://jarvis.nist.gov/}) is a collection of databases and tools to automate materials design using classical force-field, density functional theory, machine learning calculations and experiments. In particular, we obtain elastic tensor and DOS data from JARVIS-DFT database, establish the DFT workflow with JARVIS-Tools package and train the deep-learning model using ALIGNN. JARVIS-DFT is a density functional theory based database of 55645 materials with several material properties such as formation energy, bandgap with different level of theories \cite{choudhary2018computational}, solar-cell efficiency \cite{choudhary2019accelerated}, topological spin-orbit coupling spillage \cite{choudhary2021high,choudhary2019high,choudhary2020computational}, elastic tensors \cite{choudhary2018elastic}, dielectric tensors, piezoelectric tensors, infrared and Raman spectrum \cite{choudhary2020high}, electric field gradients \cite{choudhary2020density}, exfolation energies \cite{choudhary2017high} etc. with stringent DFT-convergence setup \cite{choudhary2019convergence}. Including $T_C$ in JARVIS-DFT would greatly enrich the applicability of the database and aid in guided materials design. In this work, we add the $T_C$ data in JARVIS-DFT.

\section{Results and discussion}

\subsection{First principles workflow}

A flow chart for designing conventional superconductors using BCS inspired screening (iBCS), JARVIS-Tools based DFT-Screening (J-Scr), and ALIGNN in an integrated way is shown in Fig. 1. The combination of computationally expensive DFT-based screening with  empirical rules-based screening and deep learning methods allows us to apply our combined workflow to a wider group of materials than a brute-force computational approach.
\begin{figure}[hbt!]
    \centering
    \includegraphics[trim={0 0cm 0 0cm},clip,width=0.9\textwidth]{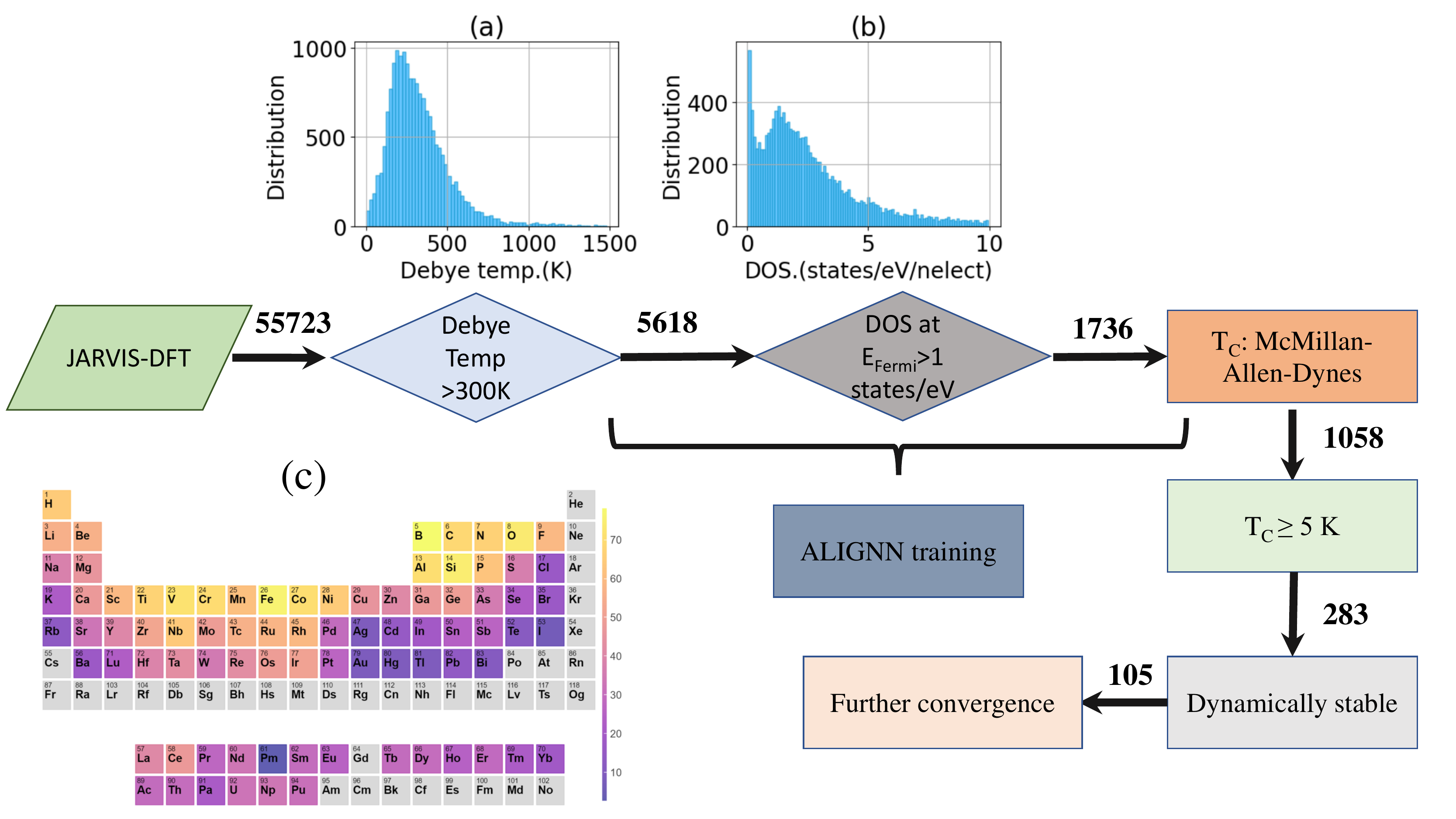}
    \caption{Schematic showing the steps involved in identifying high-$T_C$ superconductors. a) statistical distribution of Debye temperature (K) and b) statistical distribution of electronic density of states \textcolor{black}{(states/eV/total number of electrons)} at Fermi level from the JARVIS-DFT database, c) probability that compounds containing a given element have $\theta_D>$300 K. The flow chart shows the application of BCS-inspired screening, density functional theory calculations and deep-learning training.}
\end{figure}

First, we screen likely superconductors using a criteria inspired by the BCS equation for $T_C$ (Eq.1). According to this equation, high-$T_C$ materials usually have high Debye temperature ($\theta_D$), and high electronic DOS \textcolor{black}{(states/eV/total number of electrons)} at the Fermi level ($N(0)$). The JARVIS-DFT database provides DFT-based electronic DOS at the Fermi level \cite{choudhary2018computational} and the Debye temperature, as derived from the elastic tensor database\cite{choudhary2018elastic}. 

Currently, there are electronic DOS database available for 55723 materials and elastic tensors for 17419 materials (\textcolor{black}{using the v08.18.2021 version of JARVIS-DFT database at the time of writing}). We analyze the statistical distribution of $\theta_D$ and $N(0)$ 
in Fig. 1a and Fig. 1b, respectively. We observe that the Debye temperature can range up to 1500 K with high peak near 200 K (Fig. 1a). The DOS at the Fermi level can range up to 10 states/eV/Nelect with peak value around 1 (Fig. 1b). Out of 17419 materials, we find 5618 of them with $\theta_D$ greater than 300 K. Furthermore, selecting materials with electronic DOS at the Fermi level greater than 1 states/eV/Nelect, we find 1736 materials. In Fig. 1c, we present the probability that compounds containing a given element have $\theta_D>$ 300 K. We observe that light elements that form strong covalent bonds, including the $1s$, $2p$, and $3p$ elements, as well as many lighter $3d$ transition metals, tend to have high $\theta_D$, while denser transition metals and elements that form weaker bonds tend to have low $\theta_D$. \textcolor{black}{There are multiple methods to calculate Debye temperature \cite{anderson1963simplified,garrity2016first}. In this work, we primarily use the elastic tensor data, as it was available before our phonon calculations and thus available for pre-screening, but we compare to an approach based on the phonon DOS in Supplementary Figure 1.}

\begin{figure}[hbt!]
    \centering
    \includegraphics[trim={0. 0cm 0 0cm},clip,width=0.95\textwidth]{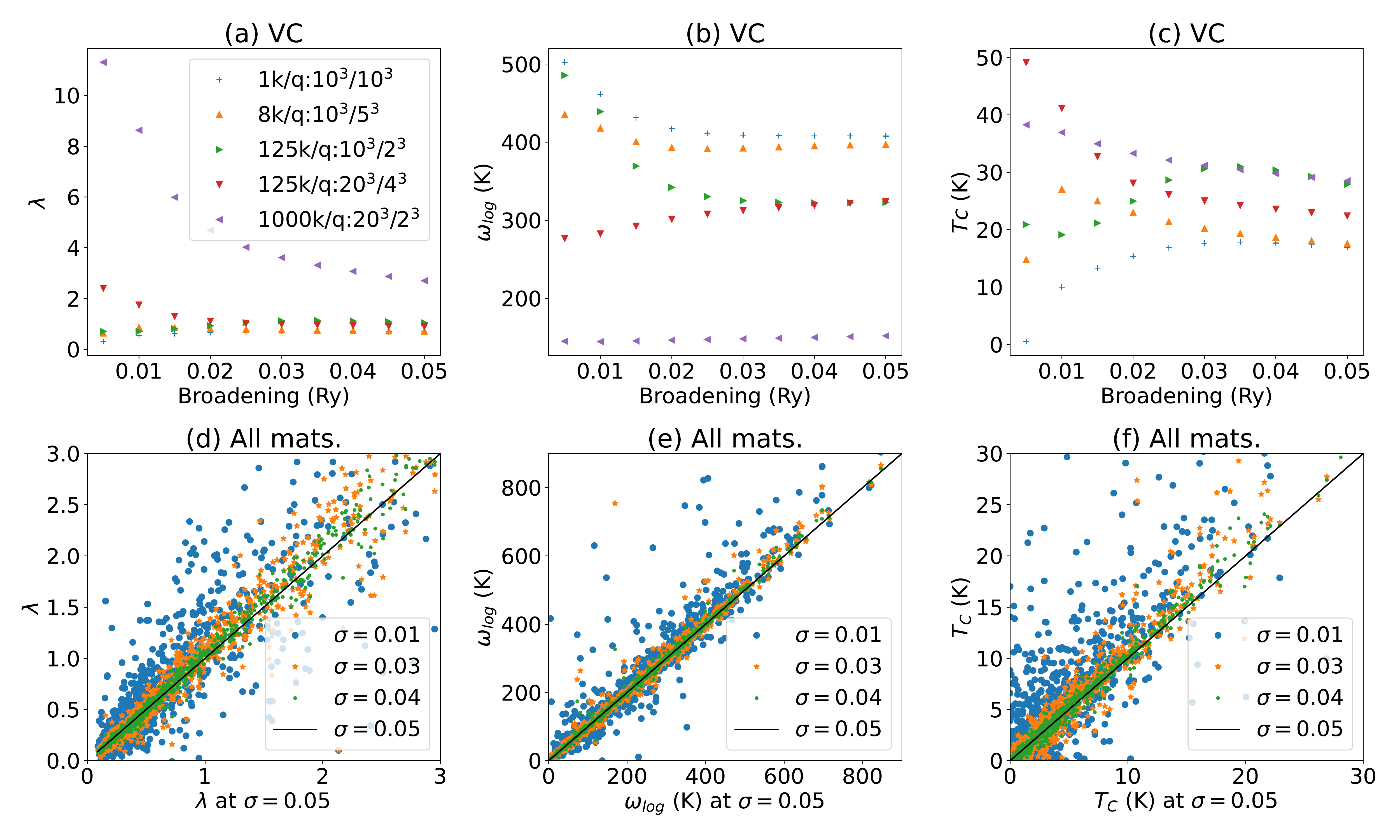}
    \caption{\textcolor{black}{Effect of different k-point and q-point selection on EPC parameters for VC. a) $\lambda$, b) $\omega_{log}$, c) $T_C$. Broadening dependent scaled d) $\lambda$, e) $\omega_{log}$ and f) $T_C$ for the all the systems in the database. The scaling was done with respect to the corresponding values at broadening of 0.05 Ry. The legend in panel a corresponds to the number of k-points and q-points, e.g. 10x10x10 k-points and 10x10x10 q-points (denoted by black star) is same as 1 k-point per q-point; and 20x20x20 k-point and 2x2x2 q-point correspond to 1000 k-point per q-point (denoted by left purple triangles). }}
\end{figure}

\begin{figure}[hbt!]
    \centering
    \includegraphics[trim={0. 0cm 0 0cm},clip,width=0.7\textwidth]{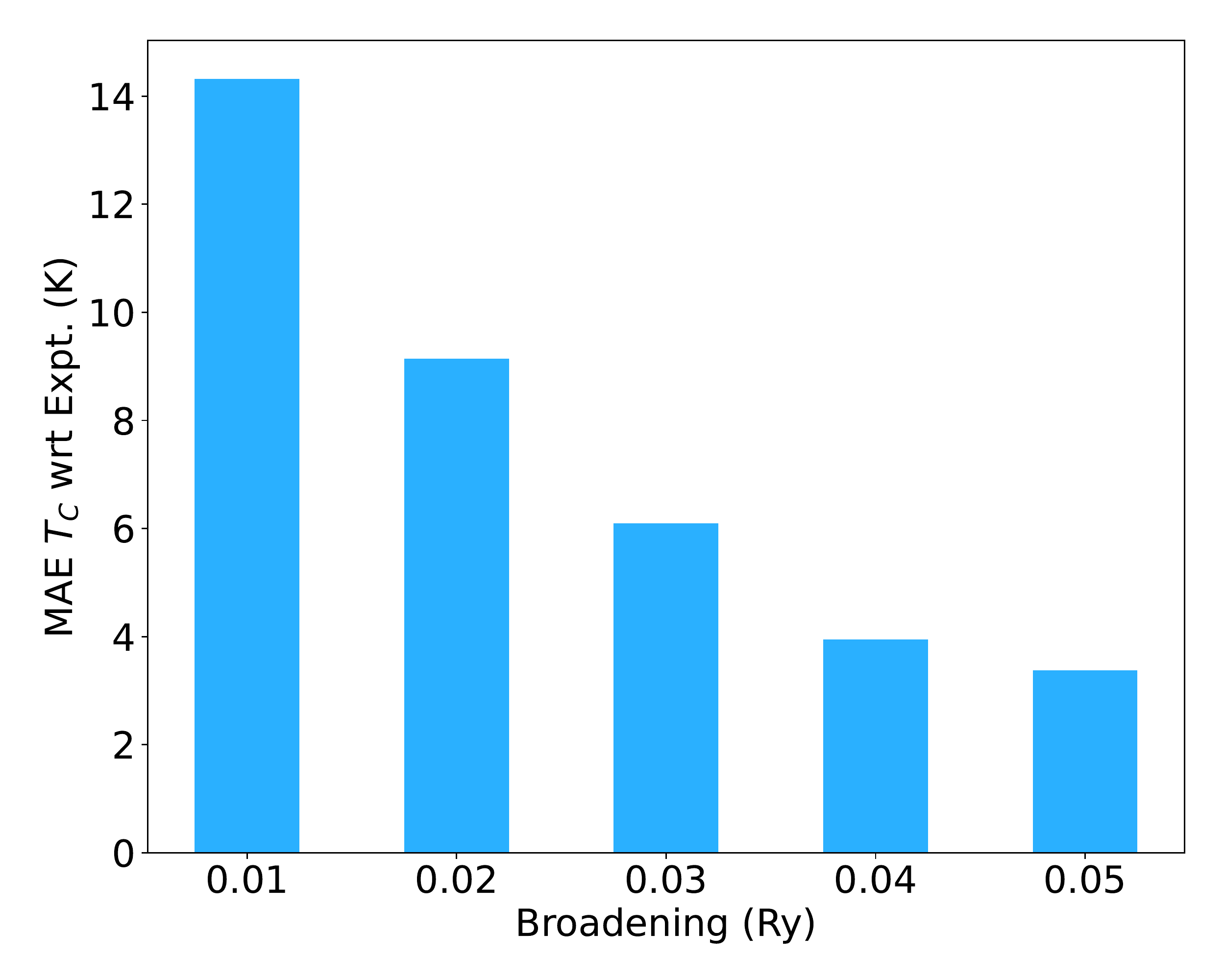}
    \caption{\label{fig:broad}\textcolor{black}{Broadening convergence with respect to experimental $T_C$ data for 14 superconductors.}}
\end{figure}

\textcolor{black}{Performing fully converged electron-phonon coupling calculations using DFT-PT is very computationally expensive, as the calculations in general require both high density $k$-point grids to sample to electronic states and high $q$-point sampling for the phonons\cite{wierzbowska2005origins}, and the number of modes to calculate at each q-point increases with the number of atoms in the primitive cell. Therefore, in the supplementary information, we perform a large number of convergence checks to understand the minimal set of convergence criteria that provide useful information in identifying possible high-$T_C$ conventional superconductors (see Supplementary Table 1-3).  We show the convergence of $\lambda$, $\omega_{log}$ and $T_C$ with respect to q-points in Supplementary Figure 2, Supplementary Figure 3 and Supplementary Figure 4 respectively. The comparison for the same for PBE \cite{perdew1996generalized} vs PBEsol are shown in Supplementary Figure 5, Supplementary Figure 6, Supplementary Figure 7 and Supplementary Figure 8 respectively. }

\textcolor{black}{In Fig. 2, we show the effect of broadening parameter on EPC and $T_C$ values. As an example of convergence results, in Fig. 2a, Fig. 2b and Fig. 2c, we show convergence with several k-point and q-point settings for the VC compound. From Fig. 2a and 2b, it is clear that denser k-points are necessary for reasonable $\lambda$ and $\omega_{log}$ values. We find that larger broadenings are necessary for lower q-point grids. Similar behavior was also found by Shipley et al. for rare-earth hydride system\cite{shipley2020stability}.  }


\textcolor{black}{In Fig. 2a, we find that though for 20x20x20 k-points and 2x2x2 q-points case, $\lambda$ and $\omega_{log}$ values (panel a and panel b) are different from other selections, the $T_C$ value is still closer to 10x10x10 and 2x2x2 k-points and q-points (panel c) respectively. We find that $T_C$ for 10x10x10 k-points and  10x10x10 q-points (black plus sign) are similar to that of 10x10x10 k-points and 5x5x5 q-points (orange upper triangle), suggesting that increasing q-points do not affect much for the 10x10x10 k-points in terms of predicting $T_C$. Similarly, for 10x10x10 k-points, 2x2x2 q-points (right green triangle); 20x20x20 k-points , 2x2x2 q-points (left purple triangle), the $T_C$ values are closer suggesting in this case increasing k-points do not change the $T_C$ much for 20x20x20 k-points. Interestingly, for the 20x20x20 k-points and 4x4x4 q-points case, the $T_C$ lies in-between the above two extremes of $T_C$. We note that though $\omega_{log}$ seem to converge and overlap for right green and downward red triangles in Fig. 2b, the $\lambda$ value is still slightly different (as from Fig. 2a) leading to large difference in $T_C$ values indicating that it harder to converge $\lambda$ that $\omega_{log}$.}

\textcolor{black}{Next, we compare the effect of broadening parameters on the EPC parameters and $T_C$ for the entire database in Fig. 2d, Fig. 2e and Fig.2f. We find relatively consistent results for broadenings of 0.03-0.05 with our relatively sparse q-point grids, while smaller broadenings lead to erratic unconverged results. In Fig.~\ref{fig:broad} we compare the effects of broadening on our prediction of $T_C$ for 14 known superconductor (see table \ref{table:expt}), and we find that higher broadening values give better agreement.}


\textcolor{black}{We investigate the effect of $\mu^*$ parameter \cite{lee1995first,giustino2017electron,marques2005ab,morel1962calculation,zarifi2018crystal,wierzbowska2005origins} in Eq. 7 on $T_C$ for several materials in Fig. 4. In Fig. 4a, we show the effect of different $\mu^*$ on 5 well-known conventional superconductors (Al, Pb, MgB$_2$, NbN, V$_3$Si). We find that as we increase the $\mu^*$ values, the $T_C$ decreases, as expected. The rate of decay somewhat is dependent on the material, but the materials remain in largely the same order as $\mu^*$ is varied. Furthermore, we compute the mean absolute error in predicting $T_C$ of 14 known superconductors in table~\ref{table:expt} using different $\mu^*$ values. We find that minimum MAE was obtained for $\mu^*=0.09$. Similarly, we compare the $T_C$ for the entire database for different $\mu^*$  in Fig. 4c. We compare these values to $\mu^*=0.09$ which is shown as a solid black line. We find that increasing $\mu^*$ decreases the $T_C$ similar to Fig. 4a. Now, we quantify, how many materials in our dataset have $T_C \geq 5K$ as we vary $\mu^*$. Relative to $\mu^*=0.09$, (denoted by dotted green line), we find that we gain 85.5 \% and 37.4 \% materials for $\mu^*$ 0.03 and 0.06 respectively. As we increase the $\mu^*$ to 0.12, 0.15 and 0.18 we lose 29 \%, 43.5 \% and 57.3 \% materials respectively.
}

Based on these tests, we find that $q$-point grids as small as $2\times2\times2$, combined with k-point grids similar to the typical grids used for self-consistent DFT total energy calculations are already useful in identifying candidate materials. Therefore, we adopt the strategy of performing an initial screening step with low convergence parameters applied to a high number of materials, which will be further refined later. During this step, we use the same $k$-points grid found during the JARVIS-DFT total energy convergence, as well as a $q$-point grid of at least $2\times2\times2$ with a broadening of 0.05 Rydberg ($\approx 0.68$ eV).

\begin{figure}[hbt!]
    \centering
    \includegraphics[trim={0 0cm 0 0cm},clip,width=0.9\textwidth]{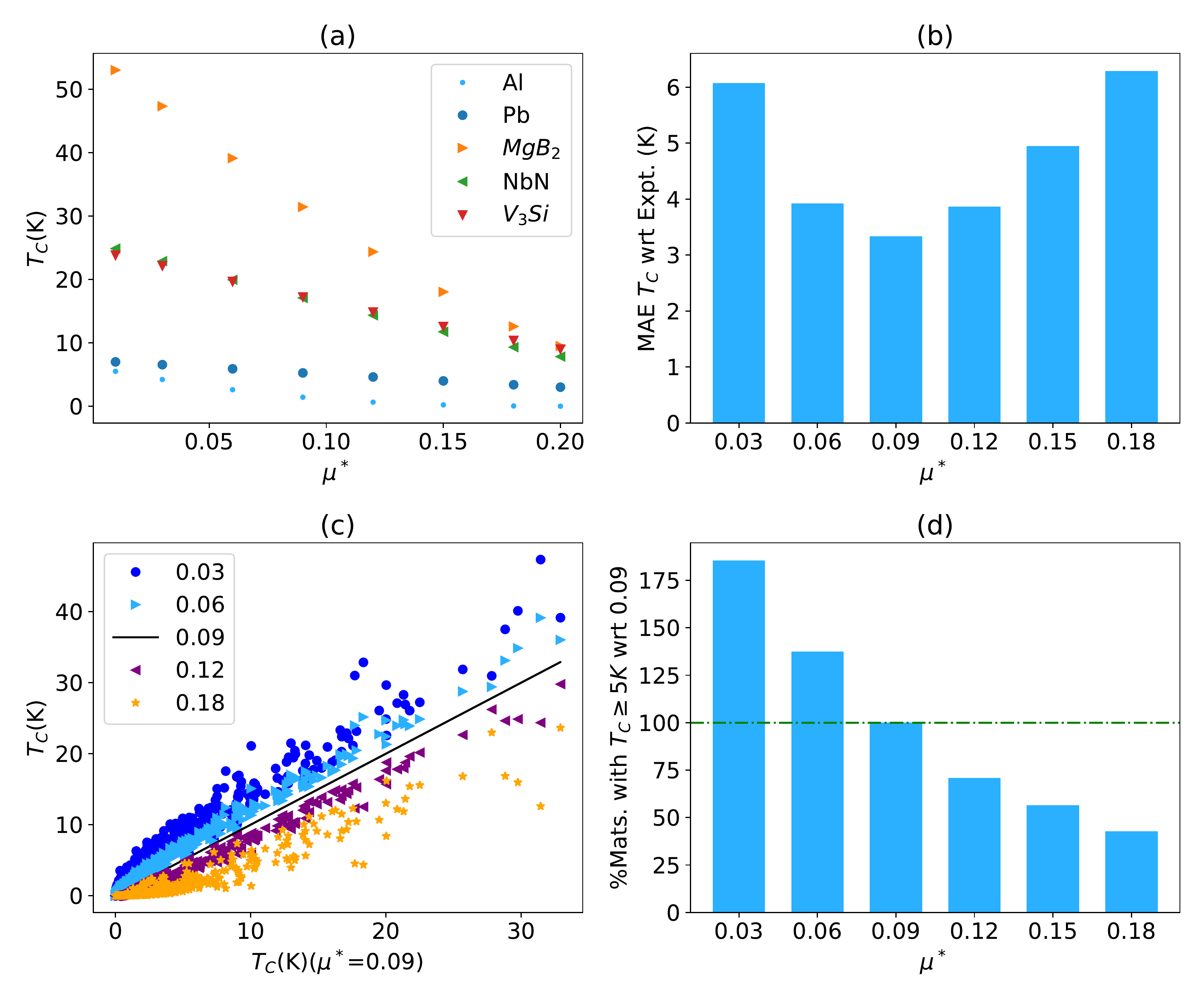}
    \caption{\label{fig:mu}\textcolor{black}{Effect of $\mu^*$ on transition temperatures. a) Change in $T_C$ as we increase $\mu^*$ for a few representative superconductors, b) change in $T_C$ with respect to $\mu^*$ 0.09 for the entire dataset, c) change in percenatge materials with $T_c$ greater equal to 5 K. Here $\mu^*$ 0.09 is shown as 100 \%. As we decrease $\mu^*$, we add materials and vice-versa.}}
\end{figure}

We compare the present DFT-screening workflow (J-Scr) with other methods such as superconducting density functional theory (SCDFT) \cite{sanna2020combining}, Lüders-Marques (LM) \cite{luders2005ab,marques2005ab} and experiments \cite{allen1975transition,roberts1976survey,webb2015superconductivity} in Table I. Note that we ignore spin-orbit coupling, spin-polarization, and \textcolor{black}{spin-fluctuation contributions \cite{heid2010effect,gibson1989evidence,mandal2014strong,gerber2017femtosecond}} during the calculations because of the additional computational cost, but these effects should be considered in follow-up investigations. Importantly, we find that for the top seven systems shown in Table I, the $T_C$ from the McMillan-Allen-Dynes formula, using $\mu^* =0.09$, from our J-Scr workflow has an excellent agreement with respect to both SCDFT and the experiments, justifying our approximations. We find that the mean absolute error (MAE) for J-Scr and experiment for all the top 14 near ambient condition superconductors is 3.3 K. Now, comparing the top 7 materials in Table I with respect to SCDFT and LM, the MAE of experiment vs J-Scr, experiment vs SCDFT, experiment vs LM, SCDFT vs J-Scr are: 1.9 K, 1.4 K, 4.9 K and 1.1 K respectively.

\begin{table}
\caption{\label{table:expt}Comparison of JARVIS-Tools screening workflow based (J-Scr) critical transition temperatures in K with experimental data (Exp) \cite{allen1975transition,roberts1976survey,webb2015superconductivity,drzazga2018characteristics,dew1979formation,tsindlekht2008linear,RevModPhysMatthias,hulm1972superconductivity,pechen1994tunneling}, SCDFT \cite{sanna2020combining} and LM methods \cite{kruglov2020superconductivity,liu2017potential}.}
\begin{tabular}{@{}lllllll@{}}
\toprule
Form. & Spg & JID&Exp&SCDFT&LM&J-Scr\\
\toprule
Al & 225 & 816 & 1.2 & 1.55 & 0.3 & 1.6 \\
Ta & 229 & 1014 & 4.5 & 5.17 &2.45 & 7.6 \\
Pb & 225 & 961 & 7.2 & 6.06 & 4.95 & 5.4 \\
Nb & 229 & 934 & 9.3 & 10.29 & 7.0 &10.7 \\
ZrN & 225 & 19679 & 10.0 & 11.6&6.12&10.0 \\
V$_3$Si & 223 & 14960 & 17 & 18.1 & 13.1&17.6 \\
MgB$_2$ & 191 & 1151 & 39 & 35.4 & 20.04 &33.0\\
V &  229 & 14837& 5.3& & &18.3\\
Nb$_3$Si &223&15938&18.0& & & 16.5\\
NbO&221&14492&1.38 & & &3.6\\
NbC&225&19889&12&&&17.1\\
NbN&221&36335&16.0&&&17.6\\
YB$_6$&221&20620&7.2& & &5.1\\
Nb$_3$Al &223 &11981 &16.8& & & 9.0\\
YH$_{10}$(250GPa)&225& &260& & &213.5\\
LaH$_{10}$(250GPa)&225& &211& && 190.0\\

\toprule
\end{tabular}
\end{table}


\begin{table}[hbt!]
\caption{JARVIS-Tools screening workflow based $T_C$ for some of the potential candidate superconductors ($T_C$), \textcolor{black}{their} chemical formula (Form.), spacegroup number (Spg), JARVIS-DFT ID (JID), Inorganic Crystal Structure Database ID (ICSD) \cite{belsky2002new} wherever available, JARVIS-DFT based formation energy ($E_{form}$ (eV/atom)) and energy above \textcolor{black}{convex hull} ($E_{hull}$ (eV)). For the complete list, refer to the supplementary information \label{tab:results}}
\begin{tabular}{@{}lllllll@{}}
\toprule
Form. & Spg & JID&ICSD&$E_{form}$&$E_{hull}$&$T_C$(K)\\
\toprule
MoN & 187 & 16897&187185   &-0.47 &0.09&33.4  \\
CaB$_2$ & 191 & 36379  &237011 & -0.25&0.09&31.0  \\
ZrN&194&13861&161885&-1.76&0.18&30.0\\
VC & 225 & 19657&619079   &-0.48&0.06&28.1 \\
V$_2$CN & 123 & 105356&-   &-0.82&0.11&26.2 \\
Mn &225&25344&41509&0.08&0.08&23.0\\
NbFeB&187&4546&-&-0.15&0.39&22.1\\
NbVC$_2$&5&102190&-&-0.46&0.08&21.9\\
ScN &225&15086&290470 &-2.15&0.0&20.8\\
LaN$_2$&2&118592&&-1.05&0.0&20.4\\
VRu&221&19694&106010 &-0.22&0.01&20.3\\
TiReN$_3$&161&36745&-&-0.68&0.10&20.0\\
B$_2$CN&51&91700&183794&-0.53&0.19&19.4\\
KB$_6$ &221 &20067 &98987& -0.09 &0.0&19.0\\
ZrMoC$_2$&166&99893&-&-0.49&0.08&17.9\\
TaB$_2$ &191&20082&30420&-0.60&0.0&17.2\\
NbS&194&18923&44992& -0.98&0.05&17.0\\
TaVC$_2$&166&101106&-&-0.54&0.05&16.3\\
TaC&187&36405&&-0.24&0.40&16.1\\
MgBH&11&120827&-&-0.03&0.11&15.5\\
CoN&216&14724&236792 &-0.02&0.0&15.0\\
NbRu$_3$C&221&8528&77216&-0.02&0.19&15.0\\
TiB$_4$Mo&191&105371&-&-0.67&0.06&14.8\\
Al$_4$CO&6&53086&-&-0.60&0.41&14.4\\
ScC&194&37100&-&-0.02&0.20&14.3\\
TaS&187&4699&52114&-0.74&0.11&14.2\\
VOs & 221 & 122961&  &-0.26 &0.0&14.1  \\
RuO$_2$& 136 & 19852 &236962 &-1.23&0.0&14.0\\
NbVCN&160&99944&-&-0.78&0.15&13.6\\
NbB$_2$ & 191 & 14726  &614908 &-0.65 &0.0&13.8   \\
CaRu$_2$N$_2$&164&108175&-&-0.32&0.10&13.3\\
ZrNbCN&160&102642&-&-1.18&0.06&13.1\\
V$_3$Al & 221 &36199&&-0.15&0.14&13.0\\
TiO&221&50092 &&-2.65&0.08&12.6\\
LiBe$_2$Ru&123&71476&&-0.14&0.08&12.5\\
V$_2$TcRu&225&41603&&-0.3&0.0&12.3\\
ScSi$_2$&191&4406&651822&-0.31&0.23&12.2\\
VTe & 221 & 122995 && 0.22 &0.50&11.3\\
AsP$_2$&191&120382&&0.45&0.45&11.2\\
Cr$_3$GaN&221&15120&626036&-0.30&0.04&10.9\\
CrH&225&18397&626118&0.06&0.06&10.7\\
TiReTc$_2$&225&39667&&-0.27&0.0&10.7\\
ZrSc&187&100313&&-0.002&0.01&10.3\\
V$_3$Pt&223&20434&649824&-0.43&0.0&10.3\\
La$_3$SnC&221&101642&-&-0.60&0.0&10.2\\
NbS&187 &20636&645320&-1.0 &0.07&10.1\\
FeN & 216 & 15718 &236789 & -0.30&0.0&10.0 \\
InP&225&17716&188692 &-0.01&0.20&10.5\\
V$_3$Ni& 223 &15777 &647028&-0.16&0.0&10.0\\
NbRu$_3$&221&16588&105224&-0.12&0.0&10.0\\
MnN&216&78284&236788 &-0.44&0.0&10.0\\
CaZrO$_2$&99&114723&&-2.76&0.30&10.0\\
ScSe&225&19974&44972 &-1.68&0.0&9.1\\


\toprule
\end{tabular}
\end{table}

This suggests that there is a close agreement between the computational methods. We also evaluate the workflow for the well-known high-pressure hydrides \cite{shipley2020stability,liu2017potential,drozdov2015conventional} such as LaH$_{10}$ and YH$_{10}$ at 250 GPa as shown in Table I. We find the MAE to be 25 K for these two systems, which is reasonable for high-throughput type screening\cite{shipley2020stability,liu2017potential,drozdov2015conventional}. Therefore, the workflow can be applied to study high-pressure superconductivity as well. However, for the present work, we analyze materials without external pressure. 

Next, we apply the J-Scr workflow on materials with number of atoms less than equal to 5, $\theta_D>$300K and $N(0)>$1 as discussed in the previous section.  As of now, we have applied J-Scr to 1058 materials. Out of these, only 626 of them are dynamically stable (i.e. no imaginary phonon modes) and 105 of them have $T_C>$5 K, which is very promising. Table II provides the chemical formula, spacegroup, JARVIS-DFT ID, Inorganic Crystal Structure Database (ICSD) ID \cite{belsky2002new}, and the OptB88vdW-based formation energy from JARVIS-DFT and predicted $T_C$ for some of the best candidates for experimental synthesis. \textcolor{black}{The JARVIS-IDs can be used to obtain more detailed information, e.g., https://www.ctcms.nist.gov/~knc6/static/JARVIS-DFT/JVASP-16897.xml for JARVIS-ID: 16897}

We find that several potential candidates are based on nitrides, borides and Vanadium containing compounds. All of these systems have high symmetry, with spacegroup numbers ranging from 99 to 225, although this may be an artifact of pre-selecting systems with few atoms. We visualize some of the crystal structures in Supplementary Figure 9. Many of these systems have hexagonal and cubic symmetry. The JARVIS-DFT identifiers for corresponding systems can be used to obtain further properties. Importantly, most of these systems have negative formation energies and energy above convex hull less than 0.5 eV/atom. We also show Fermi-surfaces of such example compounds in Fig. S9. The shape of the Fermi-surface depicts the electron motion inside a material \cite{yamase2021fermi}. In earlier works \cite{an2021superconductivity} related to MgB$_2$ type compounds, the shape of the Fermi surface helped understand the mechanism of superconductivity in such compounds.


As shown in table\ref{tab:results}, some of our highest $T_C$ candidate superconductors include MoN, VC, Mn, MnN, LaN$_2$, KB$_6$, TaC, etc. MoN in a rock-salt structure has been previously been reported with a $T_C$ as high as 30 K \cite{inumaru2008high,wang2015hardest}; however, here we propose a hexagonal form which has not been observed experimentally. Similarly, ZrN in rock-salt form is known to have $T_C$ of 10 K; however, here we propose a hexagonal form of ZrN to have a $T_C$ around 30 K. CaB$_2$ was previously theoretically predicted be a high-$T_C$ superconductor\cite{choi2009prediction} . Previous reports of synthesizing VC in 1:1 ratio has been found to be challenging by previous researchers \cite{zbasnik1973electronic}. Lanthanum nitride \cite{vaitheeswaran2002structural} in 1:1 ratio compound has been found to have superconducting properties. \textcolor{black}{Lanthanum nitride in 1:2 ratio is not characterized for the superconducting properties to the best of our knowledge.} Vanadium Ruthenium alloy with a different ratio than 1:1 has been reported earlier as superconducting with $T_C$ close to 5 K \cite{chu1971electronic}. KB$_6$ has been proposed to be a high-$T_C$ superconductor as well \cite{katsura2010possibility}. Tantalum carbide has been shown to be a potential topological superconductor with a $T_C$ of 10.3 K \cite{yan2021superconductivity}. Li \textit{et. al.} \cite{li2020superconducting} recently reported the CoN system as a potential superconductor with $T_C$ of 16 K. Nb-Ru-C alloy system has been synthesized experimentally before \cite{hong2019structural}, but no superconducting data is available to the best of our knowledge. Ta-S system with a 1:2 ratio is a superconductor \cite{sernetz1974superconductivity} with $T_C$ of 5.4 K, however, a 1:1 ratio material i.e Tantalum rich system has not been characterized for superconducting properties yet. Nb-B system has been reported earlier with a $T_C$ of around 9.8 K \cite{ren2008structural}. V-Al alloy system can attain $T_C$ of around 11.15 K as reported by Kodess et al. \cite{Kodess71}. The Sc-Si system has been realized experimentally before but their superconducting properties has not been reported\cite{gokhale1986sc}. 2D B$_2$O system was recently proposed with a $T_C$ of 10.3 K \cite{yan2020theoretical}. Cr$_3$GaN belongs to the cubic anti-perovskite family of superconductors and has been predicted to be superconductor by Tutuncu et al. \cite{tutuncu2013theoretical}. V$_3$Pt belongs to the A15 family and has been investigated for superconducting properties earlier \cite{ramakrishnan1986resistivity}. Fe-N system with 1:2 ratio was shown with a $T_C$ of 8 K \cite{chen2018novel}.

\begin{figure}[hbt!]
    \centering
    \includegraphics[trim={0. 0cm 0 0cm},clip,width=0.7\textwidth]{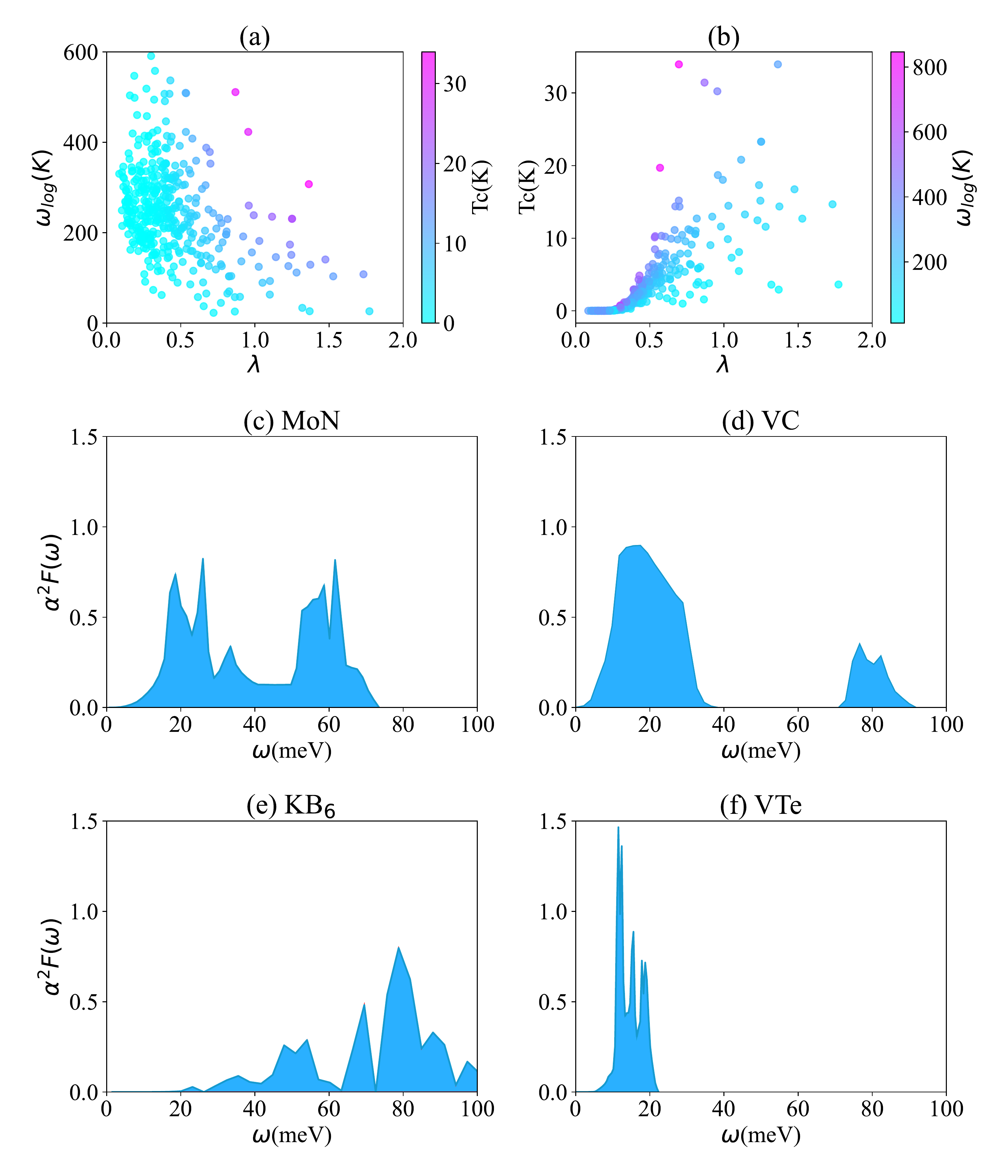}
    \caption{Relation between Electron-phonon coupling parameters and EPC function of some of the potential candidate superconductors.  a) $\omega_{log}$ vs $\lambda$, b) $T_C$ vs $\lambda$, c) MoN, d) VC, e) KB$_6$, and f) VTe. \label{fig:epc}}
\end{figure}

Next, we show the relationship between EPC parameters for the dynamically stable compounds and the Eliashberg spectral functions of some of the candidate materials in Fig.~\ref{fig:epc}. In Fig.~\ref{fig:epc}a, we find that $\lambda$ and $\omega_{log}$ to have an inverse relationship while in Fig.~\ref{fig:epc} we see that $\lambda$ and $T_C$ tend to follow a linear relationship, \textcolor{black}{ both of which are typical for BCS superconductors}. For high-$T_C$, high $\omega_{log}$ as well as high $\lambda$ is favorable as evident from the colormap. The EPC Eliashberg spectral function expresses the electron-phonon interaction in the form of a spectral density. The weighted area under the EPC function determines the $\lambda$ as well as $\omega_{log}$ parameters. 

\subsection{Machine Learning}

Next, we develop deep-learning models to accelerate both our initial BCS-inspired screening and our calculation of the electron-phonon coupling parameters. 
\begin{figure}[hbt!]
    \centering
    \includegraphics[trim={0. 0cm 0 0cm},clip,width=0.7\textwidth]{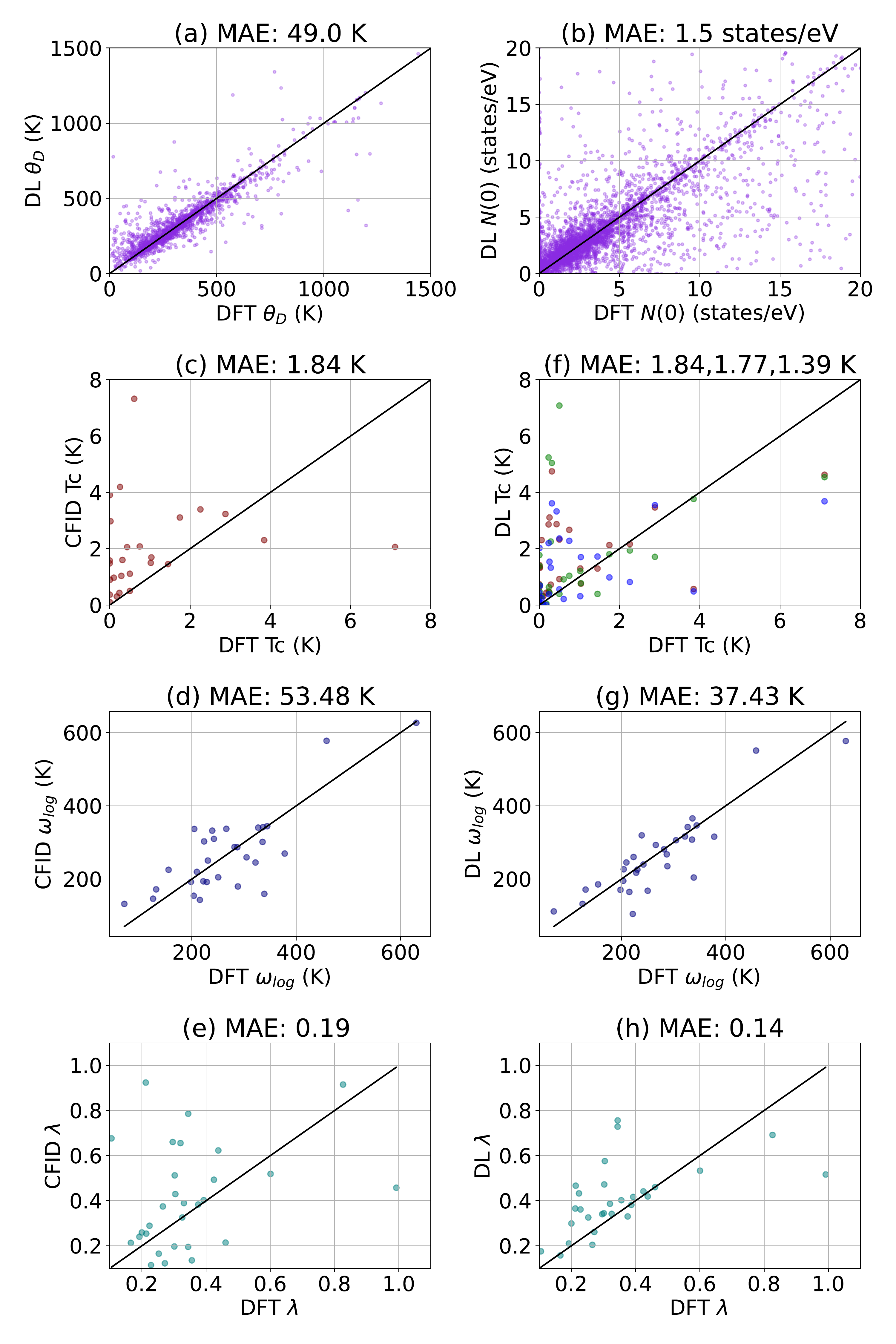}
    \caption{Atomistic line graph neural network based deep-learning (DL) regression model performance on 5 \% test set for a) Debye temperature and b) DOS.  Classical force-field descriptor (CFID) (c,d,e) and DL (f,g,h)  based regression model performance on 5 \% test set for DFT calculated transition temperature ($T_C$), EPC parameter $\omega_{log}$, and EPC parameter $\lambda$. In (f), we show performance with direct $T_C$ prediction (red color), $T_C$ prediction with direct prediction of wlog and lambda and then using eq. 7 (green color) and $T_C$ prediction with Eliashberg function and then using eq. 5-8 (black color).}
\end{figure}

While the BCS pre-screening step is much less expensive than the full EPC calculation, it still requires the DOS and $\theta_D$, which require substantial computation. Therefore, we develop  regression models to predict these properties directly from an arbitrary crystal structure, using the large datasets available in the JARVIS-DFT database. There are multiple methods to establish the structure-property relation for crystal structures, but in this work we use atomistic line graph neural network (ALIGNN) model \cite{choudhary2021atomistic}, which has been shown to outperform many well-known benchmarks for solids and explicitly capture chemical and many-body physical interactions in materials. \textcolor{black}{Our results of these models on 5 \% held test sets are shown in Fig. 6 and Table. III.} The baseline model MAE was computed by using mean of the target values in the training dataset and using it as predictions for all the materials in the 5\% test data. We observe that the mean absolute error for the Debye temperature is 49 K while that for DOS is 1.5 states/eV/Nelect. The baseline model MAE for the Debye temperature and DOS are 145.5 K and 3.62 states/eV/Nelect respectively. Using these trained models, we applied it to Crystallography Open Database (COD) \cite{gravzulis2009crystallography} with 431778 materials and with number of atoms less than 100, and pre-screened 8293 materials with high Debye temperature and DOS values. 


\begin{figure}[hbt!]
    \centering
    \includegraphics[trim={0. 0cm 0 0cm},width=0.95\textwidth]{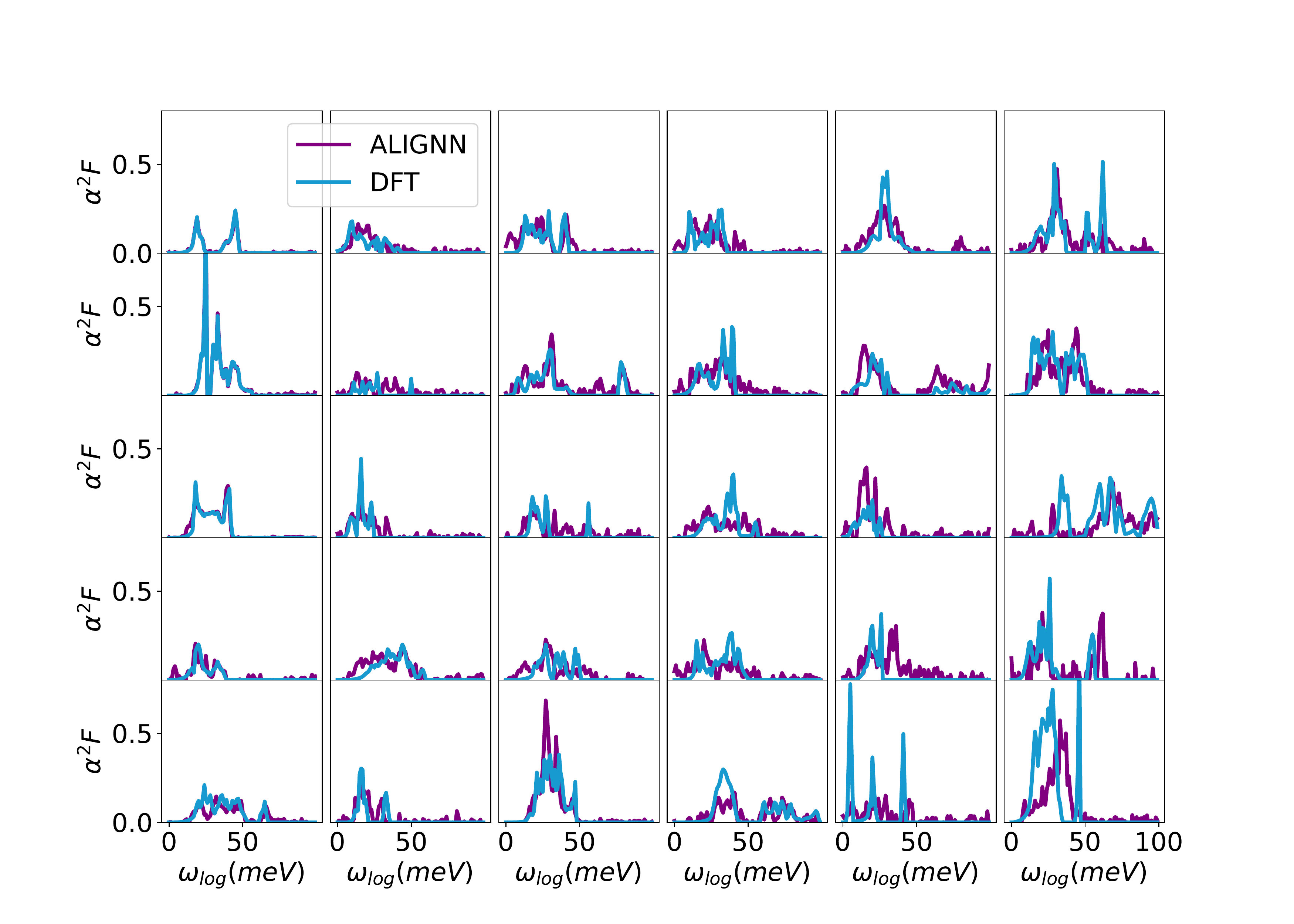}
    \caption{Prediction of Eliashberg function with ALIGNN for the 5 \% test set. ALIGNN can capture peak-positions and heights reasonably well.}
\end{figure}

Next, we develop machine-learning models to directly predict EPC parameters and $T_C$ using our database of 626 dynamically stable data-points from the j-Scr developed above. We use two methods: hand-craft descriptors (Classical force-field descriptor (CFID) \cite {choudhary2018machine}) and a deep-learning approach (atomistic line graph neural network (ALIGNN)) \cite{choudhary2021atomistic}. In particular, we train models for the McMillan-Allen-Dynes transition temperature ($T_C$) and the electron phonon coupling parameters- $\omega_{log}$ and $\lambda$.  We note that ML models usually require large datasets with sample sizes in the range of thousands, but we show preliminary but useful results already with our current dataset, which will continue to grow. 

 Classical force-field descriptor (CFID) \cite {choudhary2018machine} based performances are shown in Fig. 6c, Fig. 6d and Fig. 6e for $T_C$, $\omega_{log}$ and $\lambda$ respectively. Similarly, ALIGNN based performances are shown in Fig. 6f, Fig. 6g and Fig. 6h. For a perfect agreement, the data-points should lie on the $y = x$ line. We observe that the MAE using the CFID approach for $T_C$, $\omega_{log}$ and $\lambda$ are 1.84 K, 53.48 K and 0.19. Similarly, the MAEs for ALIGNN approach are 1.84 K, 37.43 K and 0.14 respectively.  ALIGNN outperforms CFID in predicting $\omega_{log}$, but the other performances are similar. Moreover, we notice that it is significantly easier to learn $\omega_{log}$ than $T_C$ and $\lambda$ as evident from the parity plots and table \ref{tab:dl}. In particular, the model for $\lambda$ is only slightly better than the baseline model. Using the ALIGNN models for $\omega_{log}$ and $\lambda$, and eq. 7, we predict the T$_C$ with an MAE of 1.77 K. We apply the ALIGNN TC predictor model on the COD database and find that 39595 candidate materials can have $T_C\geq10 K$ and 2161 of them have $T_C\geq15 K$.

\begin{table}
\caption{
\textcolor{black}{\label{tab:dl} DL model performance for direct property predictions. We compare the mean absolute error (MAE) of a baseline model, the MAE for the DL model on the 5\% test data, and the root mean squared error (RMSE) for the 5 \% test data. The baseline model always predicts the mean of the target values in the training dataset.}}
\begin{tabular}{@{}llll@{}}
\toprule
Property & BaselineMAE&TestMAE & TestRMSE\\
\toprule
$\theta_D$ (K)&145.5&49.0&124.62\\
$N(0)$ (states/eV/Nelect)&3.62&1.5&4.40\\
$\lambda$ & 0.20 & 0.14 & 0.24 \\
$\omega_{log}$ (K) & 76.0 & 37.43 & 50.73  \\
$T_C$ (K) & 2.72 & 1.84 & 3.31  \\
\toprule
\end{tabular}
\end{table}

As predicting $\lambda$ directly using ML models is evidently challenging, we attempt an alternate method to directly predict Eliashberg function using the ALIGNN model. We choose an energy range of 0 to 100 meV with 1 meV binsize and predict the Eliashberg functions. We show the DFT and ALIGNN based Eliashberg functions for samples in the test set in Fig. 7. \textcolor{black}{We find that the ALIGNN model does an good job of capturing most of the peaks. We calculate the $T_C$ using the ALIGNN based Eliashberg function predictions, and find the MAE is 1.39 K, which improves on our direct prediction method above by 24\%.} \textcolor{black}{Interestingly, we have observed a similar behavior for predicting phonon properties using phonon density of states as an intermediate quantity in Ref.\cite{gurunathan2022rapid}. This suggests that learning more fundamental and information-rich quantities such as Eliashberg functions can be useful for ML approaches with limited data, as compared to more direct predictions of integrated quantities.} \textcolor{black}{Unfortunately, a drawback of deep-learning models is that it is difficult to extract physical insight from their internal parameters, but we hope to investigate these ideas further in future works.}

\textcolor{black}{The above ML-exercise indicates that while $\omega_{log}$ is relatively straightforward to predict using the crystal structure, $\lambda$ and consequently $T_C$ are much more challenging. We note that deep learning models usually require larger datasets of at least ten thousand entires, which remains well beyond the data available in this work. In addition to larger datasets, these results should motivate future works on designing better descriptors and models.}


In summary, we have developed a combined high-throughput DFT and ML approach to study conventional superconductors, finding over one-hundred candidate materials with predicted $T_C \geq 5K$. We provide data from this work as well as machine learning models to help accelerate the discovery of superconductors. Because we employ a high-throughput approach to screen large databases, we employ several assumptions, but we perform significant benchmark testing as well as detailed convergence checks on particular materials found to be promising in our initial screening to verify our results. We have made our datasets and tools publicly available to enhance the reproducibility and transparency of our work. We believe that our work can be of great help to guide future computational as well as experimental efforts to discover and characterize BCS-superconductors.


\section{Methods}

In this section, we discuss details of various computational methods.

\subsection{BCS inspired screening}
According to BCS-theory\cite{bardeen1957theory} , the attractive electron-electron interaction mediated by phonons gives rise to Cooper pairs, i.e. bound states are formed by two electrons with opposite spins and momenta. BCS-theory provides the relation between the superconducting transition temperature ($T_C$), Debye temperature ($\theta_D$), electronic DOS at Fermi level $N(0)$, and electron-phonon interaction ($V$) as follows:

\begin{equation} 
T_c=1.14\theta_D \exp\bigg(-\frac{1}{N(0)V}\bigg) \label{eq:bcs}
\end{equation}

$\theta_D$ is defined as \cite{anderson1963simplified}:

\begin{equation} 
\theta_D = \frac{h}{k_B} \bigg[\frac{3n N_a \rho}{4\pi M}\bigg]^\frac{1}{3} v_m
\end{equation} 
where $h$ is Planck's constant, $k_B$ is the Boltzmann constant, $n$ is the number of atoms per formula unit, $N_A$ is Avogadro constant, $\rho$ is the crystal structure’s density, $M$ is the molar mass, and $v_m$ is the average sound velocity obtained from the elastic tensor \cite{anderson1963simplified}.


We use the  $\theta_D$ obtained from finite-difference calculations of elastic tensors\cite{choudhary2018elastic}, and $N(0)$ is obtained from the DOS at the Fermi-level as available in the JARVIS-DFT database. Currently, there are electronic DOS database available for 55723 materials and elastic tensors for 17419 materials (v08.18.2021) and the database is continuously expanding. \textcolor{black}{We normalize the electronic density of states by the total number valence electrons in the DFT calculation}. JARVIS-DFT is primarily based on Vienna Ab initio Simulation Package (VASP) \cite{kresse1996efficient,kresse1996efficiency}. software and OptB88vdW  \cite{klimevs2009chemical} functional but also contains data with different functionals and methods. In JARVIS-DFT k-points are converged with respect to total energy \cite{choudhary2019convergence}. These converged k-points are also used in subsequent electron-phonon calculations using Quantum Espresso package \cite{giannozzi2009quantum,giannozzi2020quantum} (see below).



\subsection{Density functional theory (DFT) calculations}

The superconducting properties of conventional superconductors depend on EPC. There are several methods to perform EPC calculations such as interpolated/Gaussian broadening method \cite{wierzbowska2005origins}, tetrahedron method \cite{kawamura2014improved} and other Wannier-based electron-phonon method \cite{ponce2016epw}, etc. In this work, we primarily report the interpolated method, but we also compare with the tetrahedron method (see supplementary information). We perform EPC calculations using DFT-PT\cite{baroni1987green,gonze1995perturbation} with the Quantum Espresso software package \cite{giannozzi2009quantum} and the GBRV \cite{garrity2014pseudopotentials} pseudopotentials. We report results with the PBEsol\cite{perdew2008restoring} functional in the main text, and we compare to the PBE\cite{perdew1996generalized} functional in the supplementary information. We begin with structures from the JARVIS-DFT database, and perform full relaxation using our Quantum Espresso settings. The EPC parameter is derived from spectral function ${\alpha}^2 F(\omega)$ which is calculated as follows:

\begin{equation} 
{\alpha}^2 F(\omega)=\frac{1}{2{\pi}N({\epsilon_F})}\sum_{qj}\frac{\gamma_{qj}}{\omega_{qj}}\delta(\omega-\omega_{qj})w(q)
\end{equation} 
where $\omega_{qj}$ is the mode frequency, $N({\epsilon_F})$ is the DOS at the Fermi level ${\epsilon_F}$, $\delta$ is the Dirac-delta function, $w(q)$ is the weight of the $q$ point,  $\gamma_{qj}$ is the linewidth of a phonon mode $j$ at wave vector $q$ and is given by:

\begin{equation} 
\gamma_{qj}=2\pi \omega_{qj} \sum_{nm} \int \frac{d^3k}{\Omega_{BZ}}|g_{kn,k+qm}^j|^2 \delta (\epsilon_{kn}-\epsilon_F) \delta(\epsilon_{k+qm}-\epsilon_F)
\end{equation} 
Here, the integral is over the first Brillouin zone, $\epsilon_{kn}$  and $\epsilon_{k+qm}$ are the DFT eigenvalues with wavevector $k$ and $k+q$ within the $n$th and $m$th bands respectively, $g_{kn,k+qm}^j$ is the electron-phonon matrix element. $\gamma_{qj}$ is related to the mode EPC parameter $\lambda_{qj}$ by:

\begin{equation} 
\lambda_{qj}=\frac {\gamma_{qj}}{\pi hN(\epsilon_F)\omega_{qj}^2}
\end{equation}

Now, the EPC parameter is given by:

\begin{equation} 
\lambda=2\int \frac{\alpha^2F(\omega)}{\omega}d\omega=\sum_{qj}\lambda_{qj}w(q)
\end{equation} 
with $w(q)$ as the weight of a $q$ point. 


The superconducting transition temperature, $T_C$ can then be approximated using McMillan-Allen-Dynes \cite{mcmillan1968transition} equation as follows:

\begin{equation}
T_c=\frac{\omega_{log}}{1.2}\exp\bigg[-\frac{1.04(1+\lambda)}{\lambda-\mu^*(1+0.62\lambda)}\bigg]\label{eq:mad}
\end{equation}
where
\begin{equation} 
\omega_{log}=\exp\bigg[\frac{\int d\omega \frac{\alpha^2F(\omega)}{\omega}\ln\omega}{\int d\omega \frac{\alpha^2F(\omega)}{\omega}}\bigg]
\end{equation}

\textcolor{black}{We note that several variants of McMillan-Allen-Dynes formula have been proposed previously to better account for high-$\lambda$ materials \cite{allen1975transition,allen1983theory,margine2013anisotropic,marsiglio1988iterative}, however, in this work, we opt for the simpler original McMillan-Allen-Dynes formula as our screening tool. Furthermore, we note that materials with narrow electronic bandwidth may not be well-reproduced by the formula.}
In Eq.~\ref{eq:mad}, the parameter $\mu^*$ is the effective Coulomb potential parameter. \textcolor{black}{While there are ways to calculate this parameter from first principles \cite{lee1995first}, this parameter generally varies over a relatively small range (such as 0.09 to 0.18). Following previous works \cite{lee1995first,giustino2017electron,marques2005ab,morel1962calculation,zarifi2018crystal,wierzbowska2005origins}, we take a fixed value (here $\mu^*=0.09$) when reporting our main results, and we present a discussion on the effect of varying $\mu^*$ in the later section.}







\subsection{Deep-learning training}

Deep-learning is one of the fastest developing methods especially for materials science applications \cite{choudhary2021recent}. Graph neural networks (GNN) are based on deep-learning framework because they can work with unstructured and non-Euclidean data i.e. non-grid data such as atomic structures. In this work, we use the recently developed atomistic line graph neural network (ALIGNN) \cite{choudhary2021atomistic}, which is publicly available at \url{https://github.com/usnistgov/alignn}. ALIGNN has been used to train fast and accurate models for more than 60 properties of solids and molecules with high accuracy.  We use ALIGNN to train models for the Debye temperature as well as the electronic density of states at the Fermi-level using large datasets from the JARVIS-DFT database. We also train models to predict the DFT-based transition temperatures ($T_C$) and the underlying electron-phonon coupling parameters $\omega_{log}$ and $\lambda$, using the smaller datasets computed in this work. We use both ALIGNN and the hand-crafted (Classical force-field descriptor (CFID) \cite {choudhary2018machine})-based approaches for comparison. CFID approach utilizes the LightGBM package \cite{ke2017lightgbm}. 

In ALIGNN, a crystal structure is represented as a graph using atomic elements as nodes and atomic bonds as edges. Each node in the atomistic graph is assigned 9 input node features based on its atomic species: electronegativity, group number, covalent radius, valence electrons, first ionization energy, electron affinity, block and atomic volume. The inter-atomic bond distances are used as edge features with radial basis function up to 8 $\textrm{\AA}$ cut-off. We use a periodic 12-nearest-neighbor ($N$) graph construction. This atomistic graph is then used for constructing the corresponding line graph using interatomic bond-distances as nodes and bond-angles as edge features. ALIGNN uses edge-gated graph convolution for updating nodes as well as edge features using a propagation function ($f$) for layer ($l$), atom features ($h$), and node ($i$), details of which can be found in Ref.~\cite{choudhary2021atomistic}: 

\begin{equation} 
h_i^{(l+1)}=f(h_i^l{\{h_j^l}\}_{_i})
\end{equation} 

Unlike many other conventional GNNs (\cite{Xie2018CGCNN,Schutt2018SchNet,Chen2019MEGNet}), ALIGNN uses bond-distances as well as bond-angles to distinguish atomic structures. One ALIGNN layer composes an edge-gated graph convolution on the bond graph with an edge-gated graph convolution on the line graph. The line graph convolution produces bond messages that are propagated to the atomistic graph, which further updates the bond features in combination with atom features. The ALIGNN model is implemented in PyTorch \cite{paszke2019pytorch} and deep graph library (DGL) \cite{wang2019deep}. The hyperparameters for ALIGNN are kept same as the original paper. We use the available 17419 Debye temperature data and 55645 electronic DOS at Fermi level data for model training. In addition, we use the 626 electron-phonon coupling parameters \textcolor{black}{for dynamically stable materials}, their transition temperatures and Eliashberg functions. For the Debye temperature and electronic density of states at the Fermi-level, we use a batch size of 64 for 500 epochs and 80:10:10 training-validation-testing data split, while for predicting the EPC parameters and Eliashberg functions, we use a batch size of 16, 90:5:5 split and training for 300 epochs. \textcolor{black}{The test set was never used during the training procedure}. We train use Tesla V100 SXM2 32 gigabyte Graphics processing unit (GPU), with 8 Intel Xeon E5-2698 v4 CPU cores for concurrently fetching and preprocessing batches of data during training. \textcolor{black}{Please note commercial software and hardware are identified to specify procedures. Such identification does not imply recommendation by National Institute of Standards and Technology (NIST).}

\section{Data Availability Statement}
The dataset is available at the Figshare repository: \url{https://doi.org/10.6084/m9.figshare.21370572}.

 \section{Code Availability Statement}
Software packages mentioned in the article can be found at \url{https://github.com/usnistgov/jarvis} and \url{https://github.com/usnistgov/alignn}.

\section{Acknowledgments}
K.C. and K.F.G. thank the National Institute of Standards and Technology for funding, computational, and data-management resources. K.C. thanks the computational support from XSEDE computational resources under allocation number TG-DMR 190095. Contributions from K.C. were supported by the financial assistance award 70NANB19H117 from the U.S. Department of Commerce, National Institute of Standards and Technology. 

 \section{Author Contributions}
 K.C. and K.F.G. jointly developed the workflow. K.C. carried out the high-throughput DFT calculations and developed machine learning workflow. K.C. and K.F.G. jointly analyzed the DFT data and contributed in writing the manuscript.

 \section{Competing interests}
The authors declare no competing interests.

\bibliography{sn-bibliography}


 \section{Figure captions}
Fig. 1 Schematic showing the steps involved in identifying high-$T_C$ superconductors. a) statistical distribution of Debye temperature (K) and b) statistical distribution of electronic density of states \textcolor{black}{(states/eV/total number of electrons)} at Fermi level from the JARVIS-DFT database, c) probability that compounds containing a given element have $\theta_D>$300 K. The flow chart shows the application of BCS-inspired screening, density functional theory calculations and deep-learning training.

Fig. 2 Effect of different k-point and q-point selection on EPC parameters for VC. a) $\lambda$, b) $\omega_{log}$, c) $T_C$. Broadening dependent scaled d) $\lambda$, e) $\omega_{log}$ and f) $T_C$ for the all the systems in the database. The scaling was done with respect to the corresponding values at broadening of 0.05 Ry. The legend in panel a corresponds to the number of k-points and q-points, e.g. 10x10x10 k-points and 10x10x10 q-points (denoted by black star) is same as 1 k-point per q-point; and 20x20x20 k-point and 2x2x2 q-point correspond to 1000 k-point per q-point (denoted by left purple triangles).

Fig. 3 Broadening convergence with respect to experimental $T_C$ data for 14 superconductors.

Fig. 4 Effect of $\mu^*$ on transition temperatures. a) Change in $T_C$ as we increase $\mu^*$ for a few representative superconductors, b) change in $T_C$ with respect to $\mu^*$ 0.09 for the entire dataset, c) change in percenatge materials with $T_c$ greater equal to 5 K. Here $\mu^*$ 0.09 is shown as 100 \%. As we decrease $\mu^*$, we add materials and vice-versa.

Fig. 5 Relation between Electron-phonon coupling parameters and EPC function of some of the potential candidate superconductors.  a) $\omega_{log}$ vs $\lambda$, b) $T_C$ vs $\lambda$, c) MoN, d) VC, e) KB$_6$, and f) VTe. 

Fig. 6 Atomistic line graph neural network based deep-learning (DL) regression model performance on 5 \% test set for a) Debye temperature and b) DOS.  Classical force-field descriptor (CFID) (c,d,e) and DL (f,g,h)  based regression model performance on 5 \% test set for DFT calculated transition temperature ($T_C$), EPC parameter $\omega_{log}$, and EPC parameter $\lambda$. In (f), we show performance with direct $T_C$ prediction (red color), $T_C$ prediction with direct prediction of wlog and lambda and then using eq. 7 (green color) and $T_C$ prediction with Eliashberg function and then using eq. 5-8 (black color).

Fig. 7 Prediction of Eliashberg function with ALIGNN for the 5 \% test set. ALIGNN can capture peak-positions and heights reasonably well.

\end{document}